 \documentclass[final,authoryear,5p]{elsarticle}

 \usepackage{graphicx}
\usepackage{amssymb}
\usepackage[ps2pdf,%
breaklinks=true,%
colorlinks=true,%
pdfauthor={A. Obermeier et al.},%
pdftitle={Cross-calibration of the AMS02 TRD}%
]{hyperref}

\journal{Advances in Space Research}

\begin{document}

\begin{frontmatter}


\title{Cross-calibration of the Transition Radiation Detector of AMS-02 for an Energy Measurement of Cosmic-Ray Ions}

\author{A. Obermeier}
\ead{obermeier@physik.rwth-aachen.de}
\author{M.~Korsmeier for the AMS-02 collaboration}
\address{I.~Physikalisches Institut B, RWTH Aachen University, Aachen, Germany}

\begin{abstract}

Since May 2011 the AMS-02 experiment is installed on the International Space Station and is observing cosmic radiation. It consists of several state-of-the-art sub-detectors, which redundantly measure charge and energy of traversing particles. Due to the long exposure time of AMS-02 of many years the measurement of momentum for protons and ions is limited systematically by the spatial resolution and magnetic field strength of the silicon tracker. The maximum detectable rigidity for protons is about 1.8~TV, for helium about 3.6~TV. We investigate the possibility to extend the range of the energy measurement for heavy nuclei ($Z\geq2$) with the transition radiation detector (TRD). The response function of the TRD shows a steep increase in signal from the level of ionization at a Lorentz factor $\gamma$ of about 500 to $\gamma\approx20000$, where the transition radiation signal saturates. For heavy ions the signal fluctuations in the TRD are sufficiently small to allow an energy measurement with the TRD beyond the limitations of the tracker. The energy resolution of the TRD is determined and reaches a level of about 20\% for boron ($Z=5$). After adjusting the operational parameters of the TRD a measurement of boron and carbon could be possible up to 5~TeV/nucleon.

\end{abstract}

\begin{keyword}
transition radiation \sep proportional tubes \sep space-borne detector \sep cosmic rays
\PACS 29.40.Cs \sep 98.70.Sa \sep 07.87.+v
\end{keyword}

\end{frontmatter}

\parindent=0.5 cm

\section{Introduction}

Transition radiation detectors (TRDs) have a long tradition in direct cosmic-ray measurements. They have been utilized as threshold detectors already by e.g.\ \citet{hartmann} and \citet{prince}, recently by HEAT \citep{HEAT}, and now by the AMS-02 \citep{AMSposfrac} experiment. They have been used to measure the energy of highly relativistic cosmic-ray nuclei in several other experiments: CRN \citep{CRN}, TRACER \citep{TRACER}, and CREAM\footnote{Although the exposure for CREAM was not sufficient to actually observe TR events.} \citep{CREAM}.

The light weight per area and signal response at very high energies make transition radiation detectors valuable for the direct observation of cosmic radiation on balloon-borne or space-based experiments. Transition radiation (TR) is emitted as x-ray photons \citep{cherry} by particles with Lo\-rentz factors $\gamma$ above about 1000, when they traverse boundaries of different dielectric constants. Above the threshold the TR yield increases with Lo\-rentz factor, until it reaches saturation usually above $\gamma\approx10^4$. The exact onset of TR, the shape of the response curve, and its saturation depend on the radiator materials and the detector configuration. It is an advantage of TRDs that they can be calibrated with light particles over the whole range of their response, see e.g.\ \citet{CRN}. About 100 boundary transitions are needed on average for a singly charged particle to emit one TR photon. In practice this means that a lot of boundaries are needed in a TRD (realized with foam radiators) and that an energy measurement is usually only possible for heavier nuclei that produce more signal (the TR yield scales with $Z^2$). For more information on TRDs see \citet{WakelyTR}, \citet{trdsDietrich}, and references therein.

In this paper we investigate the possibility to use the TRD of the AMS-02 experiment for energy measurements of heavy cosmic-ray nuclei. First, the AMS-02 experiment is described. Then the response curve of the transition radiation detector is determined and the energy resolution calculated.

\section{The AMS-02 experiment}

\begin{figure}
\centering
\includegraphics[width=0.75\linewidth]{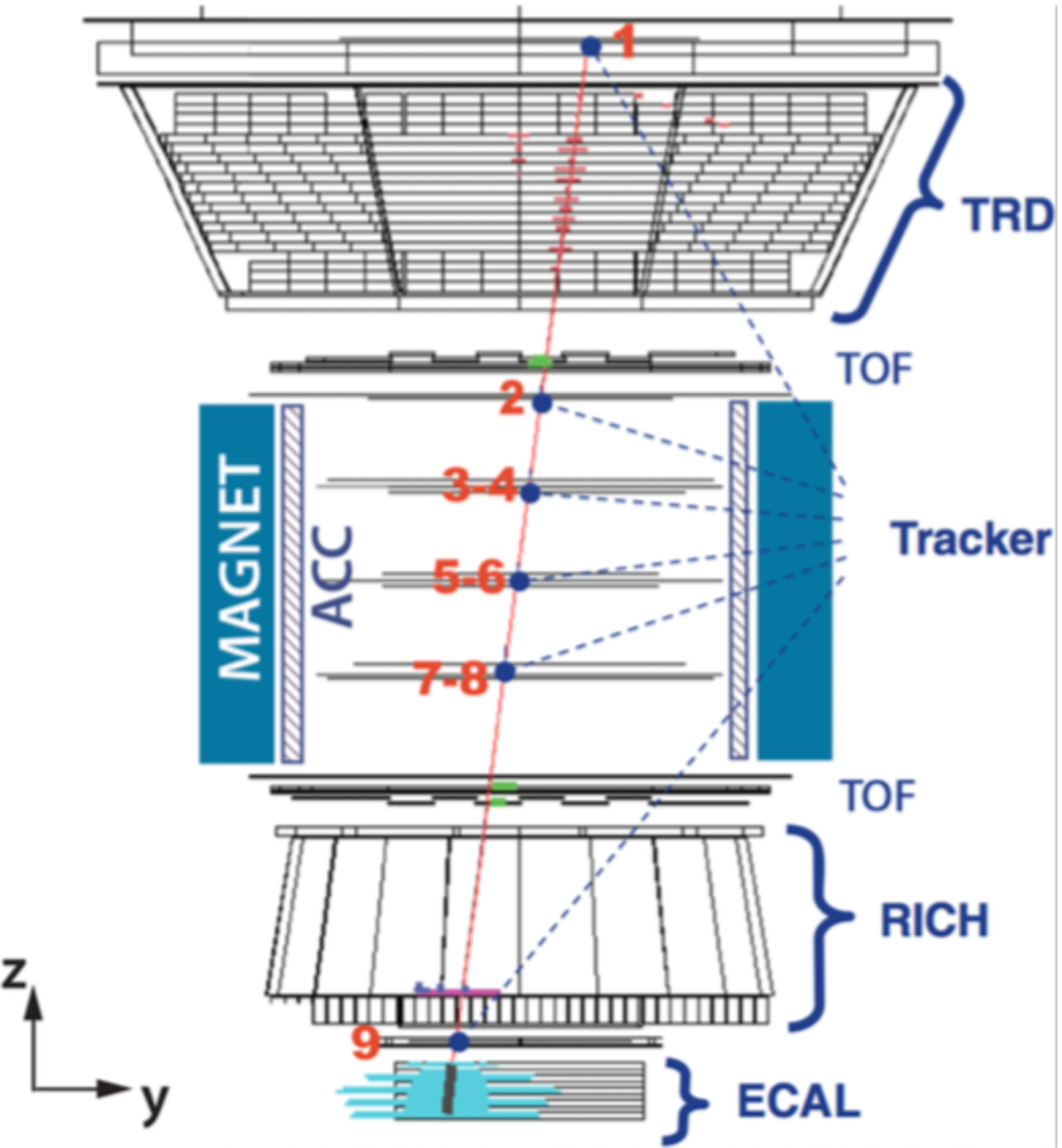}
\caption{Schematic view of the AMS-02 detector~\citep{AMSposfrac}.\label{fig01}}
\end{figure}

The AMS-02 experiment is a multi-purpose cosmic-ray detector installed on the International Space Station (ISS) since May 16th 2011. Its active detector elements consist of a silicon tracker within a strong magnetic field, a time-of-flight detector (ToF), a ring imaging Cerenkov detector, an electromagnetic calorimeter (ECAL), and a transition radiation detector (TRD). A detailed description of the detector is available by \citet{AMSposfrac} and references therein. A schematic view of the detector is shown in Figure~\ref{fig01}.

For cosmic-ray ions a measurement of rigidity is provided by the tracker. The tracker consists of nine layers of silicon sensors, of which seven layers (2 to 8) are the inner tracker located within the magnetic field. Two outer layers (1 and 9) complete the full tracker~\citep{tracker}.  Its limiting figure of merit is the maximum detectable rigidity (MDR) at which the uncertainty on the measured rigidity equals 100\%. For protons the MDR is about 1.8~TV~\citep{tracker2}, for helium it is about twice as high (3.6~TV)\footnote{This is due to a more precise determination of the intersection points of the particle track because of the larger signal helium nuclei generate in the silicon layers.}. Because of the long exposure time of AMS-02 the measurement of cosmic-ray ions will be limited by this systematic rather than by statistics. Therefore it is worthwhile to investigate the possibility to extend the energy range with the TRD.

The TRD of AMS-02 is described in detail by \citet{AMStrd2}, \citet{AMStrd}, and \citet{AMStrd3}. It consists of 5248 proportional tubes of 6 mm diameter with a maximum length of 2 m. The tubes are filled with a Xe/CO$_2$ gas mixture (90:10). They are arranged side by side in 328 modules that are mounted in 20 layers. Each layer is interleaved with a 22 mm thick fiber fleece radiator (LRP375) with a density of $0.06$ g/cm$^3$. In each layer the TRD measures the sum of the specific ionization of the particle traversing the gas and the produced transition radiation, ``dE/dx+TR''. In the energy range of the positron fraction measurement \citep{AMSposfrac}, light particles, like electrons, emit transition radiation, in contrast to heavy particles, like protons. This leads to a discrimination power between electrons and protons of up to $10^4$ for the TRD alone.

\begin{figure}
\includegraphics[width=\linewidth]{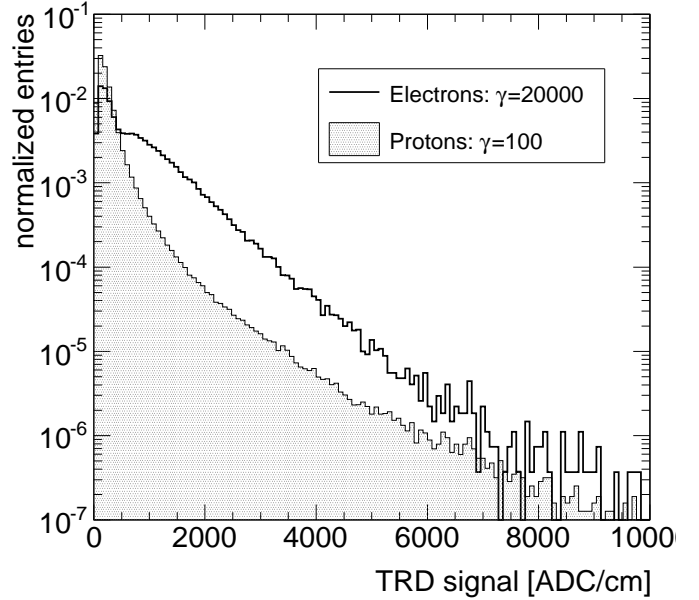}
\caption{Signal distribution in a single straw for protons (low Lo\-rentz factor, filled) and electrons (high Lo\-rentz factor, solid line).\label{fig02}}
\end{figure}

Figure~\ref{fig02} shows the pulse height spectra of a single straw for low Lo\-rentz factor (protons) and high Lo\-rentz factor (electrons) particles with $Z=1$. At low Lo\-rentz factor the signal in the TRD is purely due to specific ionization and distributed according to a Landau distribution. At high Lo\-rentz factor the signal is superimposed with a TR signal. The resulting signal distribution is greatly enhanced at high signal values compared with the Landau distribution due to specific ionization only. This difference allows the separation of protons and electrons. However, when the particle's charge is known, the signal also encodes the energy of the nucleus, as the transition radiation yield depends on the Lo\-rentz factor $\gamma$. The TRD of AMS-02 samples the signal distribution 20 times for every event. The mean signal is then clearly larger for particles with high Lo\-rentz factor than for particles with low Lo\-rentz factor.

\section{Cross-calibrating the TRD}

In order to exploit the TR emission to measure the energy of cosmic ray ions, the response curve of TRD has to be precisely known. For this purpose, the TRD was cross-calibrated with protons and electrons measured by the AMS-02 tracker and ECAL on the ISS, to determine the onset of the TR, the increase rate of the TRD signal with the Lorentz factor, and its saturation. A similar strategy has been used by the CREAM experiment~\citep{CREAMcalib}, although it could not observe particles in the TR region.

For the cross-calibration only events are considered that pass strict quality cuts, ensuring that well reconstructed events with good energy determination are used. Singly-charged particles are selected with the ToF and tracker. The tracker also determines the charge sign of every particle. Among the positively charged particles, protons are by far the most dominant species and additional cuts based on the ECAL reject a possible residual contamination of positrons. For negatively charged events, the ECAL is used to reject the small background of anti-protons from the sample in order to identify electrons.

In this study protons are used with a measured rigidity from 1 to 1000~GV in the tracker. They thus span a range of about 1 to 1100 in Lo\-rentz factor $\gamma$. Electrons are used with a measured energy in the ECAL above 0.7~GeV, which corresponds to $\gamma\approx 1400$.

\begin{figure}
\includegraphics[width=\linewidth]{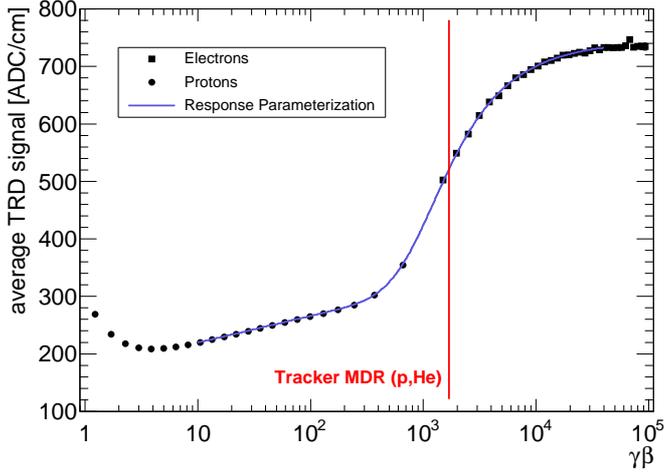}
\caption{Response curve of the TRD of AMS-02 cross-calibrated with protons and electrons. Also shown is a parameterization of the data and the MDR of the tracker\textsuperscript{\ref{foot1}}.\label{fig03}}
\end{figure}

The average TRD signal for protons and electrons is shown in Figure~\ref{fig03} as a function of Lo\-rentz factor. The signal,\\``dE/dx+TR'', is determined in each layer from the deposited energy and the corresponding track length ``dx'' determined from the precise tracking of the silicon tracker. The proton data follow an expected trend for specific ionization in gas with a minimum around $\gamma\beta\approx 3$ and a log-linear relativistic rise at higher energies. At around $\gamma\beta=1000$ the onset of TR is already evident. The increasing signal can then be traced using the electron data. The response of the TRD saturates above about $\gamma\beta=20000$. The saturated, maximum TRD signal is about 730~ADC/cm compared to 200~ADC/cm at the energy of minimum ionizing. The response curve has been found to be stable in time and stable with respect to small changes in gas mixture that occur during the operation of the detector.

Also shown in Fig.~\ref{fig03} is a parameterization of the response curve. The data have been fit by a log-linear function to which a modified Fermi function was added. The Fermi function with variable width accurately describes the onset, increase and saturation of the TR signal. The saturation of the response curve sets in beyond the tracker MDR for protons and helium\footnote{\label{foot1}For highly relativistic particles holds $\gamma\beta=Z/M\cdot R$, in which the factor of two in MDR between protons and helium is canceled by the factor of two in $Z/M$. Thus, the MDR for both species is the same in terms of $\gamma\beta$.}. It is therefore possible to extend the energy range of the AMS-02 experiment with the TRD beyond the limitations of the tracker, if the signal fluctuations are small enough to utilize the TRD for an energy measurement.

\section{Energy resolution of the TRD}

\begin{figure}
\includegraphics[width=\linewidth]{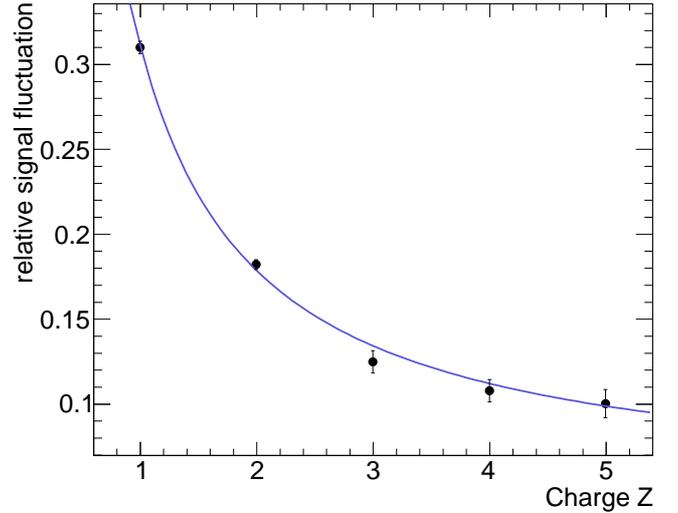}
\caption{Relative uncertainty of the average TRD signal as a function of charge at $10<\gamma<100$ for an individual measurement with up to 20 layers in the TRD. The solid line is a fit to the data according to the expected trend (see text).\label{fig04}}
\end{figure}

The intensity of transition radiation scales with the square of the traversing particle's charge. Thus, the relative fluctuations decrease for heavier nuclei proportional to $1/Z$ and the mean TRD signal for one individual event can be more precisely determined for heavier nuclei. This behavior can be seen in Figure~\ref{fig04}, in which the relative signal fluctuations are shown as a function of charge for nuclei with Lo\-rentz factors between 10 and 100.

The TRD signal at this energy range includes no transition radiation but is only due to specific ionization. This allows the analysis to include nuclei up to boron ($Z=5$) in the dynamic range of the TRD. The fluctuations of the average TRD signal are about 31\% for $Z=1$, but decrease quickly to the level of 10\% for heavier nuclei. The solid line in Fig.~\ref{fig04} represents a fit to the fluctuations $\sigma$ using (see also~\citet{swordy2}):
\begin{equation}
\sigma=\sqrt{\frac{a^2}{Z^2}+b^2},
\end{equation}
with the constants $a=0.30\pm0.05$ and $b=0.09\pm0.03$. The expected asymptotic relative signal fluctuation for heavy nuclei is therefore about 9\% and is due to uncertainties in tracking, gain calibration, and geometrical alignment of the TRD.

For protons and helium the signal fluctuations have been found to be very stable over this energy range of $\gamma$ between 10 and 100. At much higher Lo\-rentz factors, where transition radiation is present, the signal uncertainty for electrons is slightly smaller compared to the uncertainty of protons at low $\gamma$. This behavior has previously been discussed by~\citet{swordy2}. Nevertheless, the measured fluctuations of Figure~\ref{fig04} can here be used as an estimate of the detector performance.

\begin{figure}
\includegraphics[width=\linewidth]{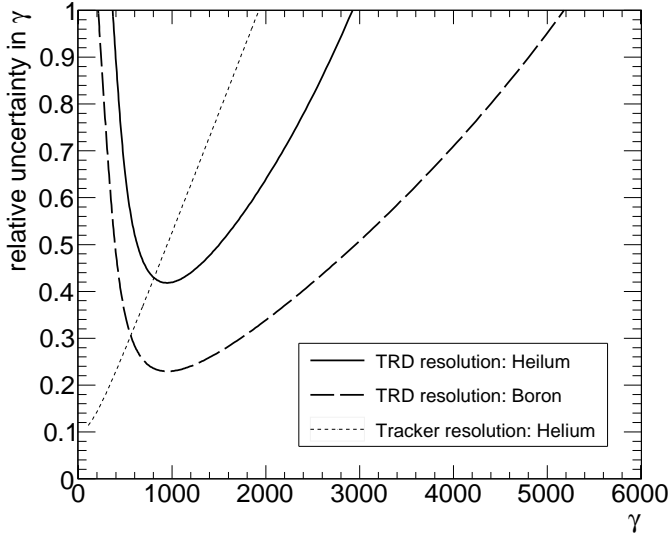}
\caption{Expected resolution for a measurement of Lo\-rentz factor with the TRD of AMS-02. Calculated for helium (solid) and boron (dashed). For comparison the tracker resolution for helium is also shown (dotted).\label{fig05}}
\end{figure}

The signal fluctuations together with the response curve determine the energy resolution that can be reached with the TRD. The calculated resolution for a measurement of the Lo\-rentz factor $\gamma$ with the TRD of AMS-02 is shown in Figure~\ref{fig05} for helium and boron. For comparison the resolution of the tracker for helium has been calculated in terms of Lo\-rentz factor and is also shown.

For helium the TRD performs better than the tracker above $\gamma\approx1000$. An energy measurement with the TRD for helium seems possible from $\gamma=500$ to 2500, above which the uncertainty reaches 100\% at about $\gamma=2800$. This could extend the energy range by a factor of 1.5 compared to the tracker. For boron, the TRD reaches a best resolution of about 20\% and a measurement may be possible in the range from $\gamma=500$ to 5000 (about 5~TeV/nucleon in energy). The tracker resolution for cosmic-ray species heavier than helium expressed in terms of Lo\-rentz factor is expected not to exceed the resolution for helium due to saturation effects in the silicon layers. Therefore, the tracker resolution for helium shown in Fig.~\ref{fig05} may serve as a approximate comparison even for boron. 

Because of the worsening TRD resolution towards lower energies a rejection of low-energy particles is needed. This can be achieved by the tracker. Straight particle tracks in the central seven layers of the tracker ensure a rigidity larger than a couple hundred~GV. This can sufficiently reject background for the TRD analysis from low energies that would otherwise spill into the high energy region and bias the measurement of cosmic-ray spectra. For helium, for which the signal fluctuations are largest and the signal distribution still has a significant Landau tail towards higher signals, this rejection of low-energy background is even more important. In this case the full tracker with all nine layers can be used to identify all low-energy spill over.

\section{Discussion and Conclusion}

The response curve of the TRD of AMS-02 determined with proton and electron data shows that a measurement of Lo\-rentz factor $\gamma$ is possible above about $\gamma=500$. For helium the accessible energy range extends to about $\gamma=2500$ or 1.5 times the MDR of the tracker. Thus, a TRD energy measurement may serve as an independent cross-check of a conventional tracker analysis. For heavier elements like boron or carbon the resolution of the TRD improves, so that a measurement up to $\gamma=5000$ may be possible, greatly extending the systematic limitations of the silicon tracker.

For a successful measurement of elemental spectra with the TRD a powerful rejection of low energy background events is imperative because of the poor energy resolution before the onset of TR. This can be achieved with the central seven layers of the tracker. Compared to a conventional analysis based on the full tracker (all nine layers) a TRD-based analysis will utilize a ten times larger geometrical acceptance, only using the TRD, ToF, and inner tracker.

Another important aspect of an energy measurement with the TRD is the dynamic range of the detector. The dynamic range of the electronics is 12~bit and at nominal operations the most probable value (MPV, not the mean value) of the signal of a minimum ionizing $Z=1$ particle is set to 60 ADC counts (this corresponds to ~100 ADC/cm and a mean value of about 200 ADC/cm due to the asymmetric signal distribution). In this configuration the saturated TR signals of high-energy helium ($Z=2$) is still well within the dynamic range. For a measurement of elements up to carbon ($Z=6$) the operation of the TRD would have to be changed so that the MPV of the signal of minimum ionizing helium corresponds to 40 ADC counts. Such a change in amplification of the initial ionization signal will change the overall normalization of the TRD response, but its shape is expected to remain stable.

\begin{figure}
\includegraphics[width=\linewidth]{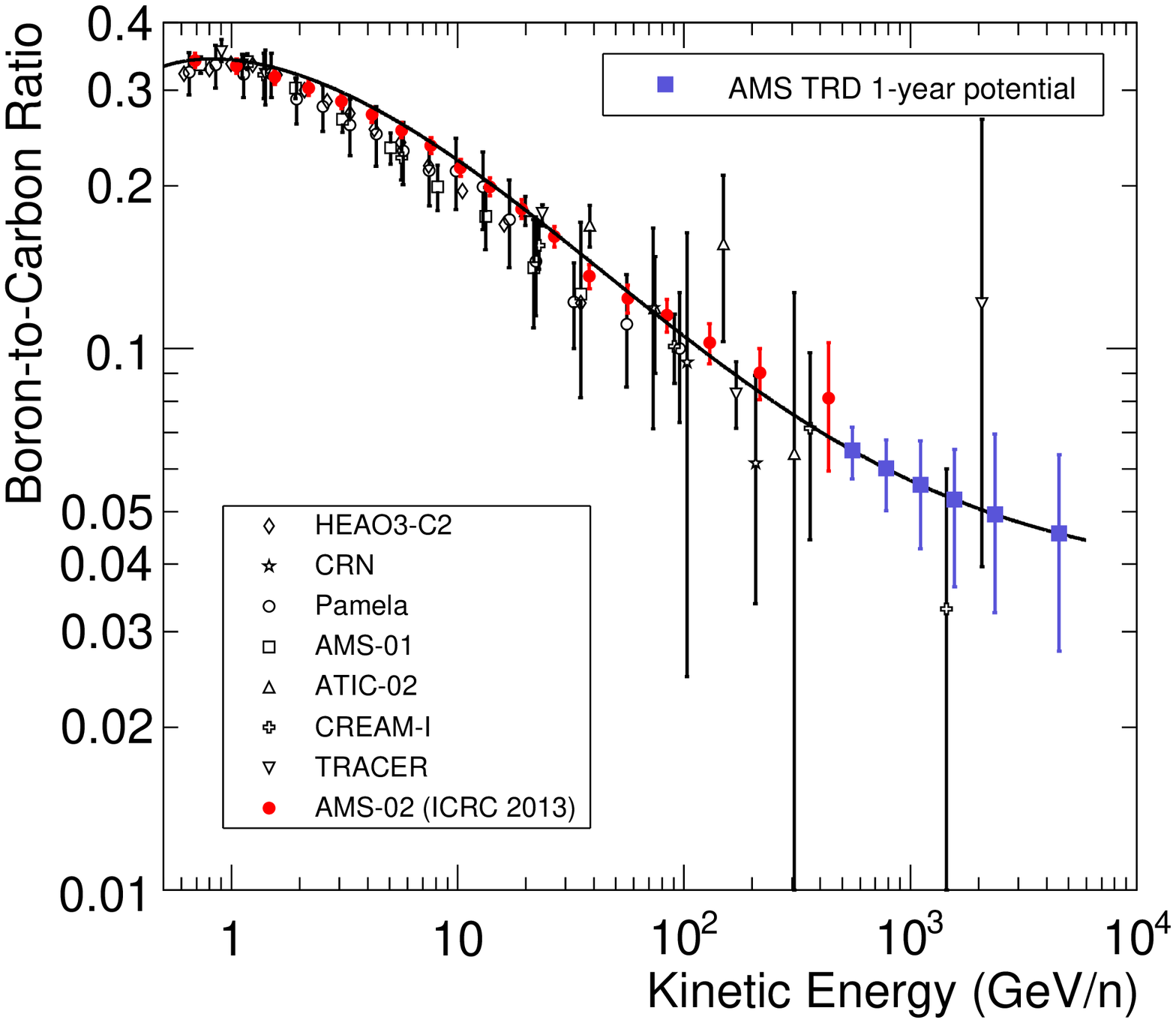}
\caption{The boron-to-carbon ratio measured by various experiments (open symbols): HEAO3-C2~\citep{heao_bc}, CRN~\citep{crn_bc}, Pamela~\citep{pamela_bc}, AMS-01~\citep{ams_bc}, ATIC-02~\citep{atic_bc}, CREAM-I~\citep{cream_bc}, and TRACER~\citep{tracer_bc}. Recent results from AMS-02~\citep{AMS02_bc} are shown (filled red circles), together with a simple model curve that is used to illustrate the possible results of a measurement of one year with the TRD of AMS-02 (filled blue squares, statistical uncertainties only). \label{fig06}}
\end{figure}

The potential of a TRD-based measurement of heavy nuclei is illustrated in Figure~\ref{fig06}. It shows the B/C ratio as measured by various experiments with the highest data points at about 1 TeV/nucleon. The expected result with statistical uncertainties of a TRD measurement with AMS-02 is added for an exposure time of one year. The measurement could reach an energy of 5~TeV/nucleon with unprecedented statistical accuracy. The expected result is based on published data on the carbon energy spectrum~\citep{tracer_bc} and the B/C ratio. A realistic live time and efficiency of the AMS-02 detector was assumed for the calculation. It should be noted that an actual measurement of cosmic-ray fluxes will of course also suffer from systematic uncertainties, for example due to corrections for interactions in the detector and unfolding of the energy scale.

\section*{Acknowledgments}

This project is supported by the German Space Agency DLR under contract number 50OO1109. The AMS experiment is grateful for the support of NASA and DOE.


\end{document}